\documentclass[rnote]{aa} 
\usepackage{graphicx}
\usepackage{natbib}                            
\usepackage{txfonts}
\begin{document}
   \title{On the morphology of the compact dust shell in the symbiotic system \object{HM\,Sagittae}\thanks{Based on observations made with the Very Large Telescope Interferometer at Paranal Observatory under programs 075.D-0484, 077.D-0216, and 079.D-0213}$^{,}$\thanks{Reduced visibilities and differential phases are available in electronic form at the CDS via anonymous ftp to {\tt cdsarc.u-strasbg.fr (130.79.128.5)} or via {\tt http://cdsweb.u-strasbg.fr/cgi-bin/qcat?J/A+A/?/?}}}

\titlerunning{The dust shell of \object{HM\,Sge}}

   \author{S. Sacuto
          \inst{1}
          \and
          O. Chesneau\inst{2}
          }

   \institute{Institute for Astronomy (IfA), University of Vienna,
              T\"urkenschanzstrasse 17, A-1180 Vienna, Austria\\
              \email{stephane.sacuto@univie.ac.at}
         \and
             UMR 6525 H. Fizeau, Univ. Nice Sophia Antipolis, CNRS, Observatoire de la C\^{o}te d'Azur, Av. Copernic, F-06130 Grasse, France\\
             \email{Olivier.Chesneau@obs-azur.fr}
             }

   \date{Received September 15, 1996; accepted March 16, 1997}
 
  \abstract
   {The symbiotic system \object{HM\,Sagittae} consists of a Mira star and a secondary White Dwarf component. The dust content of the system was severely affected by the nova outburst in 1975, which is still ongoing. The capabilities of optical interferometry operating in the mid-IR allow us to investigate the current geometry of the dust envelope.}
   {We test our previous spectro-interferometric study of this system with new interferometric configurations, increasing the $uv$ coverage and allowing us to ascertain the appearance of the source between 8 and 13~$\mu$m.}
   {We used the MIDI instrument of the VLTI with the unit telescopes (UTs) and auxiliary telescopes (ATs) providing baselines oriented from PA=42$^{\circ}$ to 127$^{\circ}$. The data are interpreted by means of an elliptical Gaussian model and the spherical radiative transfer code {\tt DUSTY}.}
   {\rm We demonstrate that the data can be reproduced well by an optically thick dust shell of amorphous silicate, typical of those encountered around Mira stars, whose measured dimension increases from 8 to 13~$\mu$m. We confirm that the envelope is more extended in a direction perpendicular to the binary axis. The level of elongation increases with wavelength in contrast to our claim in a previous study.}
   {The wider $uv$ coverage allows us to deepen our previous investigations of the close circumstellar structure of this object.}


   \maketitle
%

\section{Introduction}
\label{intro}

   \object{HM\,Sge} is a D-type (dust dominated IR) symbiotic system that erupted as a symbiotic nova in 1975 \citep{dokuchaeva76}, evolving from a 17$^{\rm th}$ to a 11$^{\rm th}$ magnitude star with a rich emission-line spectrum. The cool component is a Mira star and the hot component is a White Dwarf (WD), which had escaped detection being enshrouded in the dense Mira envelope before this dramatic event. \\
Previous works by \citet{schild01} and \citet{bogdanov01} were published on \object{HM\,Sge} the same year, evaluating the parameters of the dust in the system by fitting the ISO/SWS data using the {\tt DUSTY} radiative transfer code \citep{ivezic99}. Both groups used single and double component models but did not reach identical conclusions. \citet{schild01} favored a 2-shell model, which accounted for the effect of the White Dwarf on the dust, whereas \citet{bogdanov01} proposed a single-shell model. Since the extensions of the dusty shell differ from one model to another, mid-infrared interferometry data enabled \citet{sacuto07} to determine the most suitable of the proposed models. The authors showed that the morphology of the dust was accounted for well by the single shell model of \citet{schild01}, which included a compact and optically thick circumstellar environment, typical of a Mira star. However, because of its high optical depth, the model is inconsistent with the presence of non-absorbed emission of the Mira in the near-IR, and \citet{sacuto07} indicated that this issue could be solved by assuming a more complex spatial distribution of dust. We attempt to confirm the morphology of the dusty envelope using observations acquired with five new interferometric baselines secured in 2007.


\section{Observations}
\label{obs}

The Very Large Telescope Interferometer (VLTI) of ESO's Paranal Observatory was used with MIDI, the MID-infrared Interferometric recombiner \citep{leinert03}. MIDI combines the light of two telescopes and provides spectrally resolved visibilities in the N band atmospheric window.\\

The first observations of \object{HM\,Sge} were conducted in 2005 and 2006 with the VLT unit telescopes (UTs) UT2, UT3, and UT4, providing projected baselines and projected angles, in the range of 32-59 meters and 42$^{\circ}$ to 105$^{\circ}$ respectively (see the first two parts of Table~\ref{journal}). An interpretation of these data is described in \citet{sacuto07}. \\

The acquisition images enabled us to confirm that the mid-IR source was unresolved by the data from the individual UT. This implied that most of the mid-infrared flux originated in the inner 300 mas of the source, corresponding to the Airy pattern of the UTs and defining the Field Of View (FOV) of the interferometric observations. This also means that the flux detected within the 1.1$\arcsec$ FOV of the VLTI auxiliary telescopes (ATs), originates in this inner region. Therefore, the visibilities extracted from the ATs and the UTs are fully consistent with each other.\\ Complementary observations were completed in 2007 with the ATs A0, D0, G1, and H0, providing projected baselines and projected angles, in the range of 64-89 meters and 72$^{\circ}$ to 127$^{\circ}$ respectively (see the third part of Table~\ref{journal}). With longer projected baselines, these configurations allow us to probe smaller circumstellar regions. The new projected angles provide additional constraints about the morphology of the dust shell. \\

Table~\ref{journal} presents the journal of interferometric observations. The calibrators, \object{HD188512} (G8IV, diam=1.98$\pm$0.02\,mas), \object{HD187642} (A7V, diam=3.22$\pm$0.01\,mas), \object{HD187076} (M2II, diam=8.05$\pm$0.13\,mas), \object{HD167618} (M3.5III, diam=11.33$\pm$0.04\,mas), and \object{HD206778} (K2Ib, diam=8.38$\pm$0.09\,mas)\footnote{MIDI calibrator database}, were observed immediately before or after each science target observation. The third part of the table gathers the additional observations of the object obtained in April, May, and July 2007.\\

\begin{table}[h]
\caption{\label{journal}Journal of all available MIDI observations. The calibrators used to calibrate the visibilities are given below the science target. The phase of the Mira ($\varphi_{\rm Mira}$) during the observations is indicated. The configuration used for the observations is given. The length and position angle of the projected baseline are also indicated.}
\begin{center}
\begin{tabular}{cccccc}
\hline
\hline
{\tiny Star} & {\tiny UT date \& Time} & {\tiny $\varphi_{\rm Mira}$} & {\tiny Config.} & {\tiny Base[m]} & {\tiny PA[deg]} \\
\hline
{\tiny \object{HM\,Sge}} & {\tiny 2005-07-23 06:48:42} & {\tiny 0.75} & {\tiny U2-U3} & {\tiny 46.5} & {\tiny 44} \\
{\tiny \object{HD188512}} & {\tiny 2005-07-23 07:12:10} & {\tiny \ldots} & {\tiny -} & {\tiny \ldots} & {\tiny \ldots} \\
{\tiny \object{HM\,Sge}} & {\tiny 2005-07-24 02:41:11} & {\tiny 0.75} & {\tiny U2-U3} & {\tiny 32.1} & {\tiny 42} \\
{\tiny \object{HD188512}} & {\tiny 2005-07-24 02:58:55} & {\tiny \ldots} & {\tiny -} & {\tiny \ldots} & {\tiny \ldots} \\
{\tiny \object{HM\,Sge}} & {\tiny 2005-07-24 03:42:18} & {\tiny 0.75} & {\tiny U2-U3} & {\tiny 37.8} & {\tiny 47} \\
{\tiny \object{HD188512}} & {\tiny 2005-07-24 04:03:14} & {\tiny \ldots} & {\tiny -} & {\tiny \ldots} & {\tiny \ldots} \\
{\tiny \object{HM\,Sge}} & {\tiny 2005-07-24 06:14:49} & {\tiny 0.75} & {\tiny U2-U3} & {\tiny 46.2} & {\tiny 47} \\
{\tiny \object{HD188512}} & {\tiny 2005-07-24 06:37:58} & {\tiny \ldots} & {\tiny -} & {\tiny \ldots} & {\tiny \ldots} 
\\\hline\hline
{\tiny \object{HM\,Sge}} & {\tiny 2006-05-17 09:04:02} & {\tiny 0.31} & {\tiny U3-U4} & {\tiny 59.3} & {\tiny 105} \\
{\tiny \object{HD187642}} & {\tiny 2006-05-17 09:27:14} & {\tiny \ldots} & {\tiny -} & {\tiny \ldots} & {\tiny \ldots} \\
{\tiny \object{HM\,Sge}} & {\tiny 2006-06-11 09:05:09} & {\tiny 0.36} & {\tiny U3-U4} & {\tiny 46.8} & {\tiny 101} \\
{\tiny \object{HD187642}} & {\tiny 2006-06-11 08:42:52} & {\tiny \ldots} & {\tiny -} & {\tiny \ldots} & {\tiny \ldots} 
\\\hline\hline
{\tiny \object{HM\,Sge}} & {\tiny 2007-04-04 09:39:22} & {\tiny 0.93} & {\tiny A0-G1} & {\tiny 89} & {\tiny 118} \\
{\tiny \object{HD187076}} & {\tiny 2007-04-04 09:06:17} & {\tiny \ldots} & {\tiny -} & {\tiny \ldots} & {\tiny \ldots} \\
{\tiny \object{HM\,Sge}} & {\tiny 2007-05-08 08:52:55} & {\tiny 0.99} & {\tiny G1-D0} & {\tiny 66} & {\tiny 127} \\
{\tiny \object{HD167618}} & {\tiny 2007-05-08 10:16:28} & {\tiny \ldots} & {\tiny -} & {\tiny \ldots} & {\tiny \ldots} \\
{\tiny \object{HM\,Sge}} & {\tiny 2007-05-10 08:34:58} & {\tiny 0.99} & {\tiny H0-D0} & {\tiny 60} & {\tiny 78} \\
{\tiny \object{HD187076}} & {\tiny 2007-05-10 08:55:17} & {\tiny \ldots} & {\tiny -} & {\tiny \ldots} & {\tiny \ldots} \\
{\tiny \object{HM\,Sge}} & {\tiny 2007-05-16 09:48:33} & {\tiny 0.00} & {\tiny H0-D0} & {\tiny 64} & {\tiny 72} \\
{\tiny \object{HD206778}} & {\tiny 2007-05-16 09:20:11} & {\tiny \ldots} & {\tiny -} & {\tiny \ldots} & {\tiny \ldots} \\
{\tiny \object{HM\,Sge}} & {\tiny 2007-07-30 05:54:54} & {\tiny 0.15} & {\tiny A0-G1} & {\tiny 65} & {\tiny 107} \\
{\tiny \object{HD206778}} & {\tiny 2007-07-30 05:23:12} & {\tiny \ldots} & {\tiny -} & {\tiny \ldots} & {\tiny \ldots} \\
\hline
\end{tabular}
\end{center}
\end{table}

Chopped acquisition images were recorded (f=2Hz, 2000 frames, 4~ms per frame) to ensure the accurate acquisition of the target. The acquisition filter was an N-band filter. Photometry was obtained before and after the interferometric observations with the HIGH-SENS mode of MIDI, using the prism that provides a spectral dispersion of about 30. The data reduction software packages\footnote{\tt{http://www.mpia-hd.mpg.de/MIDISOFT/, http://www.strw.leidenuniv.nl/$\sim$nevec/MIDI/}} MIA and EWS \citep{jaffe04} were used to prepare the spectra and visibilities \citep{chesneau05,ratzka07}. MIA is based on a power spectrum analysis and uses a fast Fourier transformation (FFT) to calculate the Fourier amplitude of the fringe packets, while EWS uses a shift-and-add algorithm in the complex plane, averaging appropriately modified individual exposures (dispersed channeled spectra) to obtain the complex visibility. \\ 
An extensive study of the fluctuations of the instrumental transfer function was carried out in all periods. The rms scatter of these calibration measurements was $\sim$0.05 between 8~$\mu$m and 13~$\mu$m, corresponding to a relative error of 10\%.\\ 

All visibility and differential phase data as well as all the characteristics of the observations are available from the CDS (Centre de Donn\'{e}es astronomiques de Strasbourg); all data products are stored in the FITS-based, optical-interferometry data-exchange-format (OI-FITS) described by \citet{pauls05}.\\

\begin{figure}[tbp]
\begin{center}
\includegraphics[width=8.0cm]{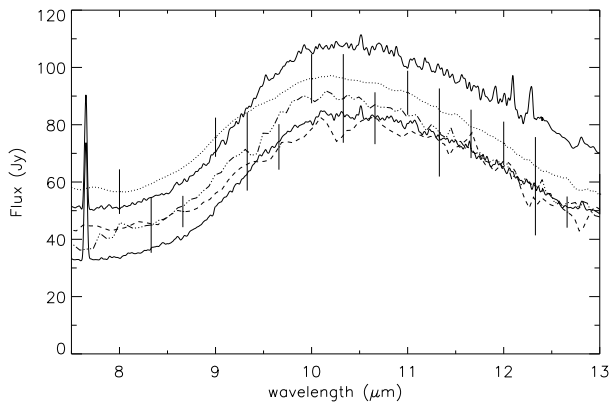}
\end{center}
\caption{MIDI and ISO/SWS flux of the source in the mid-infrared. Dotted line with $\sim$10\% error bars is the 2005 MIDI flux from the UTs ($\phi$$\sim$0.75); dashed line with $\sim$10\% error bars is the 2006 MIDI flux from the UTs ($\phi$$\sim$0.30); dashed dot dot dot line with $\sim$25\% error bars is the 2007 MIDI flux from the ATs ($\phi$$\sim$0.00). The two solid lines are the ISO/SWS spectra (upper one: 1996, $\phi$=0.65; bottom one: 1997, $\phi$=0.08)}
\label{Flux_HMSge_05_06_07}
\end{figure}

In Fig.~\ref{Flux_HMSge_05_06_07}, the MIDI spectra for each period (2005-2006-2007) is compared with the ISO/SWS spectra \citep{sloan03} acquired at two different phases of the Mira. The level of flux received by MIDI is equivalent to that of ISO, implying that most of the N flux is concentrated within the 300 mas beam of a 8 m telescope. It also means that the level of emission has probably not changed significantly since the ISO observations (1996-1997), keeping in mind a typical error for the absolute SWS flux calibration of 15 to 30\% \citep{schaeidt96}. For physical and technical reasons, the flux error bars of the photometry from the UTs are below the 10\% level, whereas the accuracy of the photometry extracted from the ATs is much worse with an error bar of 25\% (see Fig.~\ref{Flux_HMSge_05_06_07}). The error levels are estimated from the standard deviation of the flux-calibrated spectra determined from each night and each telescope, $i.e$ typically 4 to 8 spectra. As a consequence, it is difficult to differentiate between photometric variations in one phase from another. The number of measurements acquired in each period is also limited. Therefore, we decided in the following to continue the spectro-interferometric study of the star neglecting its variability.

\section{Radiative transfer modeling}
\label{sp_int_model}

The spectro-interferometric observations are interpreted in terms of spherical models generated by the {\tt DUSTY} package \citep{ivezic99}. We first fit the spectrophotometric data of the star, then we determine the corresponding synthetic spectrally-dispersed visibility profiles throughout the N band (8-13$\mu$m) and compare them with the MIDI visibilities for each baseline. \\

\subsection{The compact single-dust-shell model}
\label{comp_shell}
\begin{figure*}[tbp]
\begin{center}
\includegraphics[width=6.0cm]{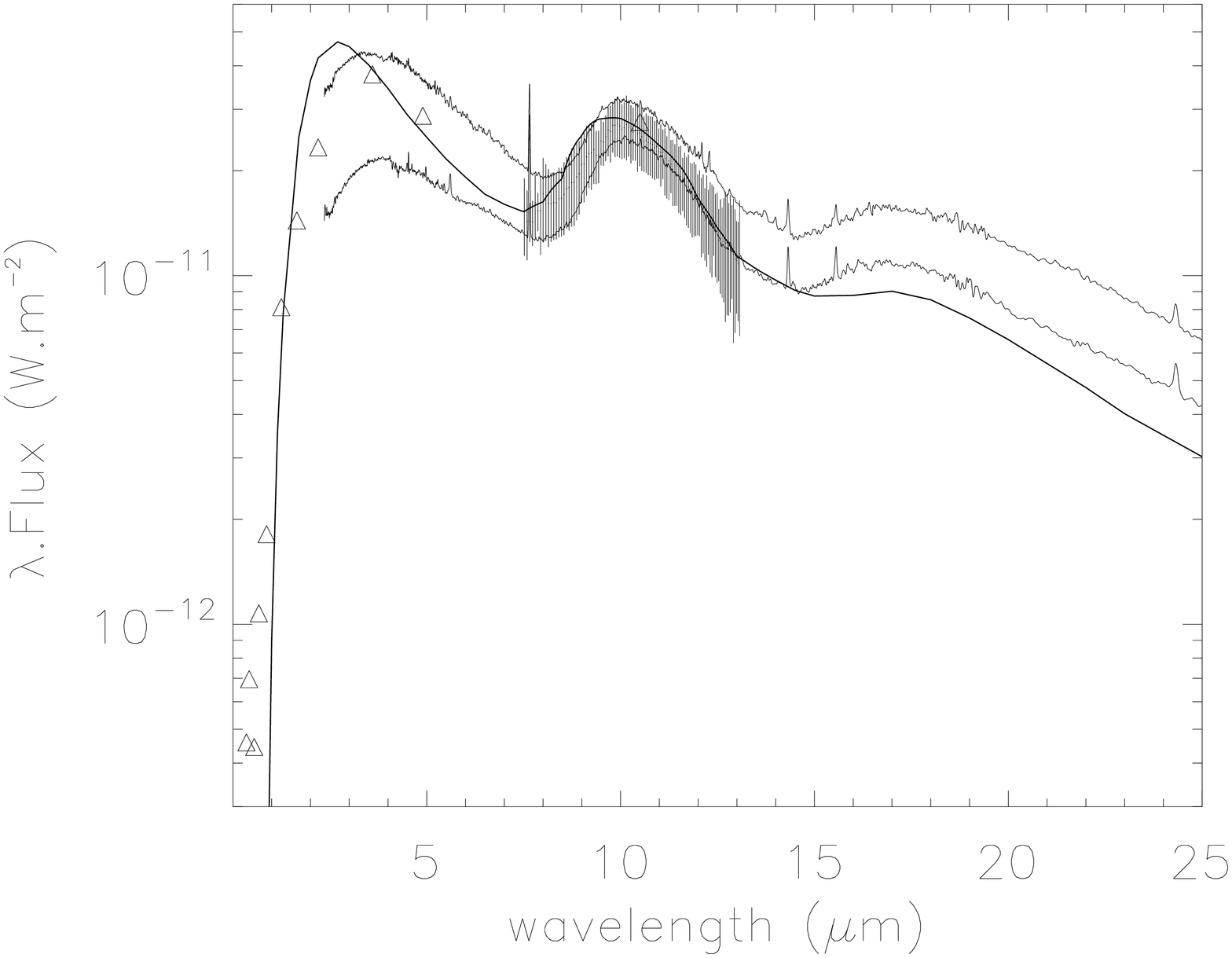}
\hspace{1.3cm}
\includegraphics[width=6.0cm]{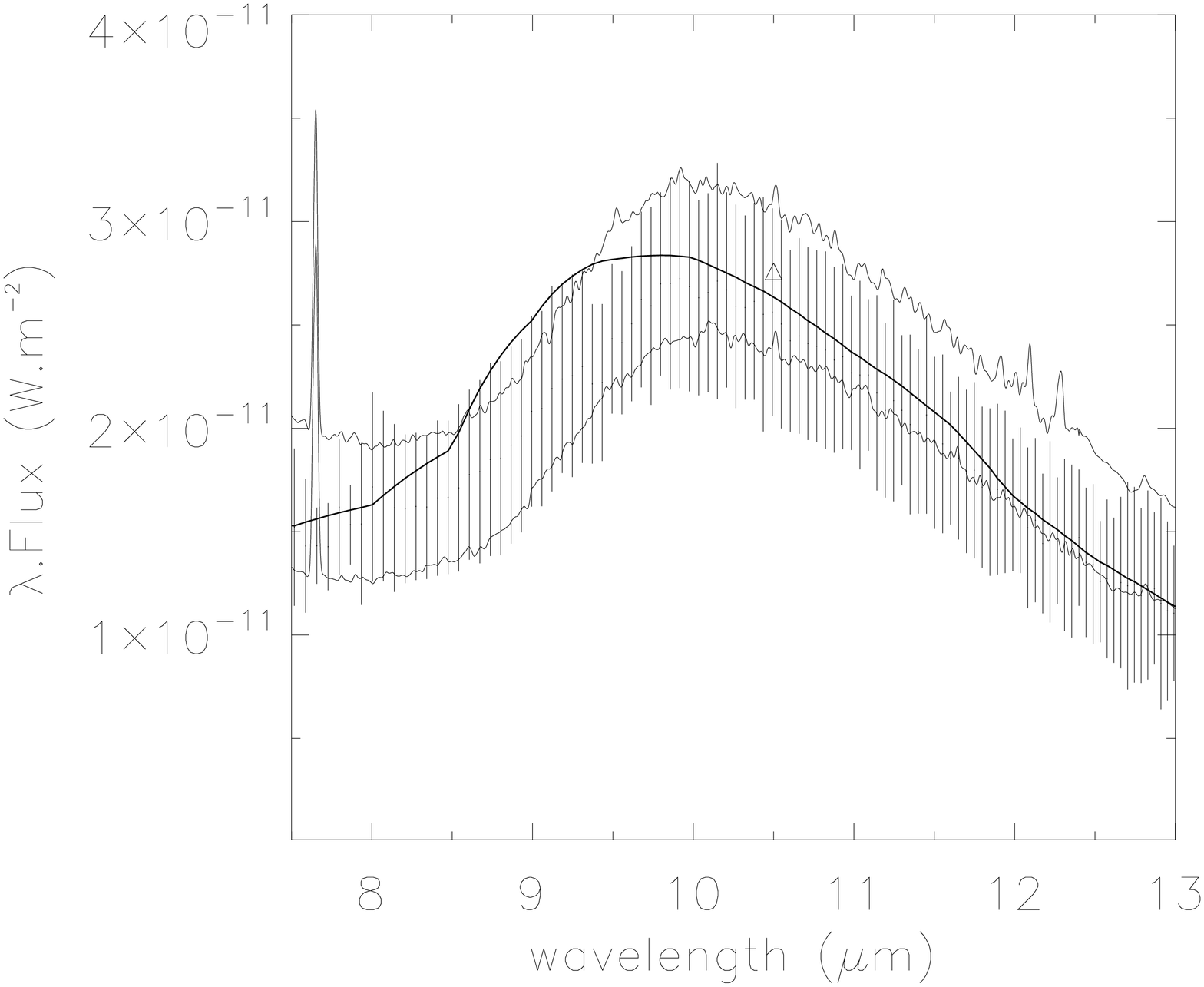}
\begin{minipage}[c]{16cm}
\vspace{0.5cm}
\centering
\includegraphics[width=13.0cm]{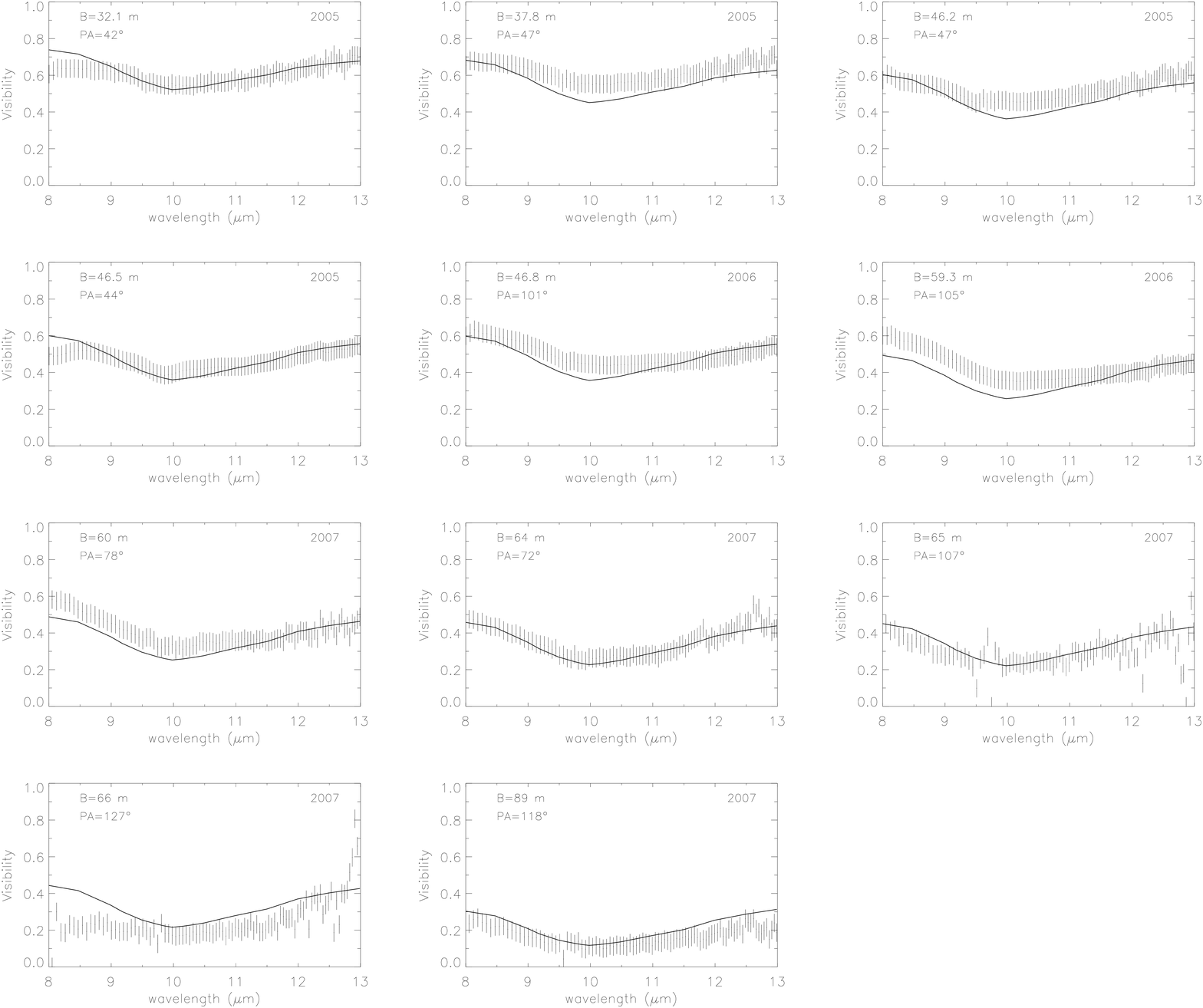}
\centering
\end{minipage}
\end{center}
\caption{\textit{Top-left}: Best-fit model of the single dust shell spectrum (thin line) supperimposed on the ATs MIDI flux (error bars), the 2 ISO/SWS spectra (upper thick line: 1996, $\phi$=0.65, and bottom thick line: 1997, $\phi$=0.08), and the photometric data (open triangles) with (\textit{U/B/V/R/J/H/K/L/M/N}) data from \citet{yudin94} and \textit{I} band from \citet{chochol04}. \textit{Top-right}: close-up view of the best-fit model to the dust feature. \textit{Bottom}: the corresponding model visibility (solid line) superimposed on the MIDI visibilities (error bars) for the eleven projected baselines.}
\label{plot_dusty_model}
\end{figure*}

The best-fit models obtained in our previous work (see Sect.~5 of \citet{sacuto07}) involved only a single shell component around the Mira. 
The modeling was based on the simultaneous fit of the ISO/SWS spectrometric \citep{sloan03} and interferometric data for periods 2005 and 2006 (see the first two part of Table~\ref{journal}). Different models were obtained: one involving pure amorphous silicates, the other including carbon material in the dusty layers. We tested these models upon the new MIDI interferometric measurements (see the third part of Table~\ref{journal}) and also some near-infrared photometric data not used in the first study.\\ 
Near-infrared photometric data provide an additional constraint of the radiation originating in the Mira star, whereas the interferometric data provide the necessary high angular resolution to probe the internal layers and to derive information about the morphology of the structure.\\
A least square fitting minimization was performed taking into account spectrophotometric and interferometric data. This minimization was completed for a large grid of parameter values by varying the effective temperature of the star from 2800 to 3200~K ($\Delta$T$_{\rm eff}$=100~K), the inner boundary temperature from 1400 to 1800~K ($\Delta$T$_{\rm in}$=100~K), the density power-law coefficient from 1.8 to 2.4 ($\Delta$p=0.1), and the 10$\mu$m optical depth from 1.5 to 2.5 ($\Delta$$\tau_{\rm 10 \mu m}$=0.05).\\
Figure~\ref{plot_dusty_model} shows the best-fit model allowing a simultaneous fit of spectrophotometric and interferometric data of the star. Table~\ref{tab_hm_sge_dusty_1shell} presents the parameters of \object{HM\,Sge} deduced from this fit.

\begin{table}[tbp]
\caption{Parameters for \object{HM\,Sge} deduced from the best-fit {\tt DUSTY} model to the spectrophotometric and interferometric data.}
\label{tab_hm_sge_dusty_1shell}
\begin{minipage}[h]{10cm}
\begin{tabular}
[c]{cc} \hline\hline
Parameter & Value \\ \hline
Effective temperature (K)  & 3000 \\ 
Central star diameter (mas) & 1.6 \\
Luminosity (L$_{\odot}$) & 5000 \\
Distance (kpc) & 1.5 \\
Shell inner radius: $\varepsilon_{\rm in}$ (mas)  & 3.6 \\
Inner boundary temperature (K)  & 1600 \\
Grain chemical composition  & 100$\%$ Sil-Ow\footnotemark[1]\\
Density power law coefficient  & 2.2 \\
Grain size distribution  & MRN\footnotemark[2] \\
Geometrical thickness ($\varepsilon_{\rm in}$)& 1000 \\
Visual optical depth & 25.2 \\
10 $\mu$m optical depth & 2.15 \\
\hline
\end{tabular}
\end{minipage}
\footnotemark[1]{Sil-Ow stands for 'warm' silicates of \citet{ossenkopf92}.}\\
\footnotemark[2]{grain size distribution as described by \citet{mathis77}.}\\
\end{table}

The results do not differ significantly from the parameters of the best-fit model deduced in the previous work of \citet{sacuto07}, and reproduce well the near-infrared photometric data and the new interferometric data of the star. However, the new data enable us to exclude the model including carbon dust proposed to improve the fit. This assumption, involving a large fraction of carbon, is not physically justified given the spectral type of the dust-forming source, an M-type Mira star. In the absence of any clear signature of other dust species and given the spatial complexity of the source, it is extremely difficult to use the dust composition as a parameter for improving the fit.\\

The interpretation by means of a single dust shell surrounding the Mira star was first rejected by \citet{schild01} because of its high opacity, which was inconsistent with the detection of the Mira photometric cycle in the near-infrared. However, interferometric data imply that a simple dust shell is the most suitable solution. \\
With the help of these new observations, the increase of the $uv$ coverage allows us now to investigate the evidence of departure of the circumstellar environment from spherical symmetry.

\section{Morphological interpretation}
\label{morp}

\begin{figure}[tbp]
\begin{center}
\includegraphics[width=8.0cm]{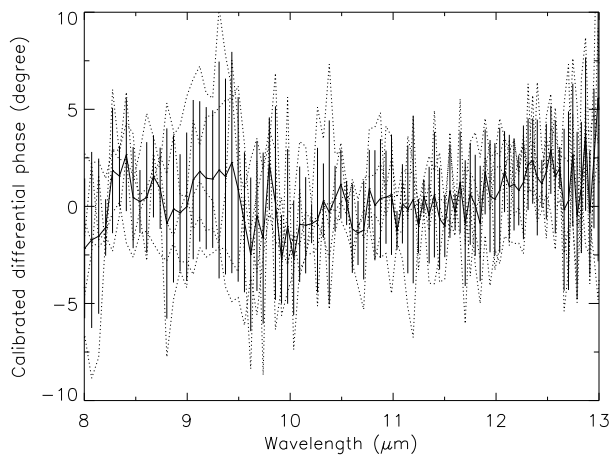}
\end{center}
\caption{Calibrated differential phase of the object evaluated from all the determined ATs projected baselines (dotted lines). Error bars correspond to the rms of the differential phase averaged over all the projected baselines. The bold solid line is the central value of these error bars.}
\label{fig_diff_phase}
\end{figure}

The phase shifts or differential phases is information extracted from the difference between the phase at a given wavelength and the mean phase determined in the full N-band region. It provides partial information (depending on the $uv$ coverage) about the asymmetry of the source, in addition to the information contained in the visibility amplitudes. As an example, strong phase signals were detected in sources harboring dusty disks \citep{deroo07,ohnaka08}.\\
Previous results of differential phases obtained by \citet{sacuto07} indicated no obvious signature of any asymmetry in the close environment of the star. Due to the low rms ($<$\,$\pm$5$^\circ$) of the calibrated differential phases averaged over all the projected baselines, new interferometric configurations confirm the global centrosymmetry of the object in the vicinity of the source (see Fig.~\ref{fig_diff_phase}). This allows us to justify in the following the utilization of a centrosymmetric geometrical model.\\

The geometrical constraints obtained from the data presented in \citet{sacuto07} show that the circumstellar environment of the star is elongated perpendicularly to the binary axis (PA=130$^\circ$ NE; \citealt{eyres01}), in the direction of the Raman line polarization \citep{schmid00}. This information was obtained by comparing the half width at half maximum of the equivalent one-dimensional Gaussian distribution from various projected baseline position angles. With the new baselines, we can perform a more robust analysis using a two-dimensional fitting in the Fourier space by means of an elliptical Gaussian flux distribution, appropriate for studying deviations from spherical symmetry.

\subsection{Elliptical Gaussian model}
\label{ellip_gauss}

The use of purely geometrical models can help to constrain the object morphology. Because the circumstellar envelope is optically thick ($\tau_{\rm 0.55\mu m}=25.2$; see Table~\ref{tab_hm_sge_dusty_1shell}), a Gaussian brightness distribution centered on the star is a natural approximation of the flux distribution in the plane of the sky.\\

The visibility of the elliptical Gaussian model can be expressed by \citep{tycner04}

\begin{equation}
\label{vis_ellip_gaus}
V(u^{\prime},v^{\prime})=exp\left[\frac{-\pi^{2} \theta_{\rm mj}^{2} \left(r^{2} u^{\prime^{2}} + v^{\prime^{2}}\right) } {4\,ln2} \right]
\end{equation}
where $\theta_{\rm mj}$ is the angular size FWHM of the major axis, $r$ is the axial ratio of the minor to the major axis, and ($u^{\prime}$,$v^{\prime}$) are the modified spatial frequencies. These frequencies can be expressed in terms of the spatial frequencies ($u$,$v$) by using a coordinate transformation 

\begin{eqnarray}
u^{\prime}=u \cos \phi - v \sin \phi \\
v^{\prime}=u \sin \phi + v \cos \phi
\end{eqnarray}
where $u=\frac{B_{\rm p}}{\lambda} \cos{\theta_{\rm p}}$ and $v=\frac{B_{\rm p}}{\lambda} \sin{\theta_{\rm p}}$, where $B_{\rm p}$ and $\theta_{\rm p}$ are the projected baseline and the projected angle on the sky respectively, and $\lambda$ is the observed wavelength. Finally, $\phi$ is the position angle (measured east from north) of the major axis.\\

\begin{figure}[tbp]
\begin{center}
\includegraphics[width=8.0cm]{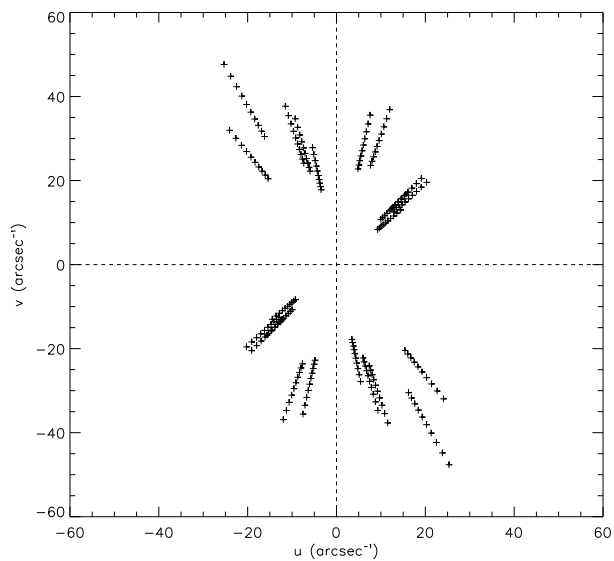}
\end{center}
\caption{Spectrally-dispersed $uv$ coverage of the 2005-2006-2007 observations from 8 to 12.5 $\mu$m.}
\label{uv_cov}
\end{figure}

\begin{figure*}[tbp]
\begin{center}
\includegraphics[width=18.0cm]{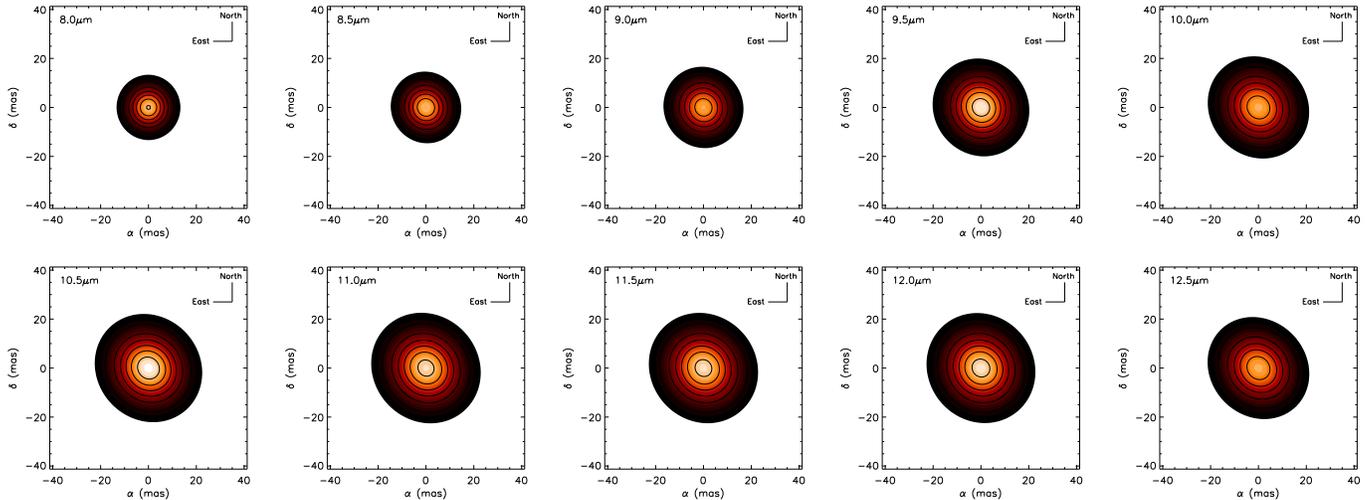}
\end{center}
\caption{Brightness distribution of the best-fit elliptical Gaussian model of \object{HM\,Sge} from 8 to 12.5 $\mu$m (north is up and east is left).}
\label{fig_morp_uv}
\end{figure*}

\begin{figure*}[tbp]
\begin{center}
\includegraphics[width=6.0cm]{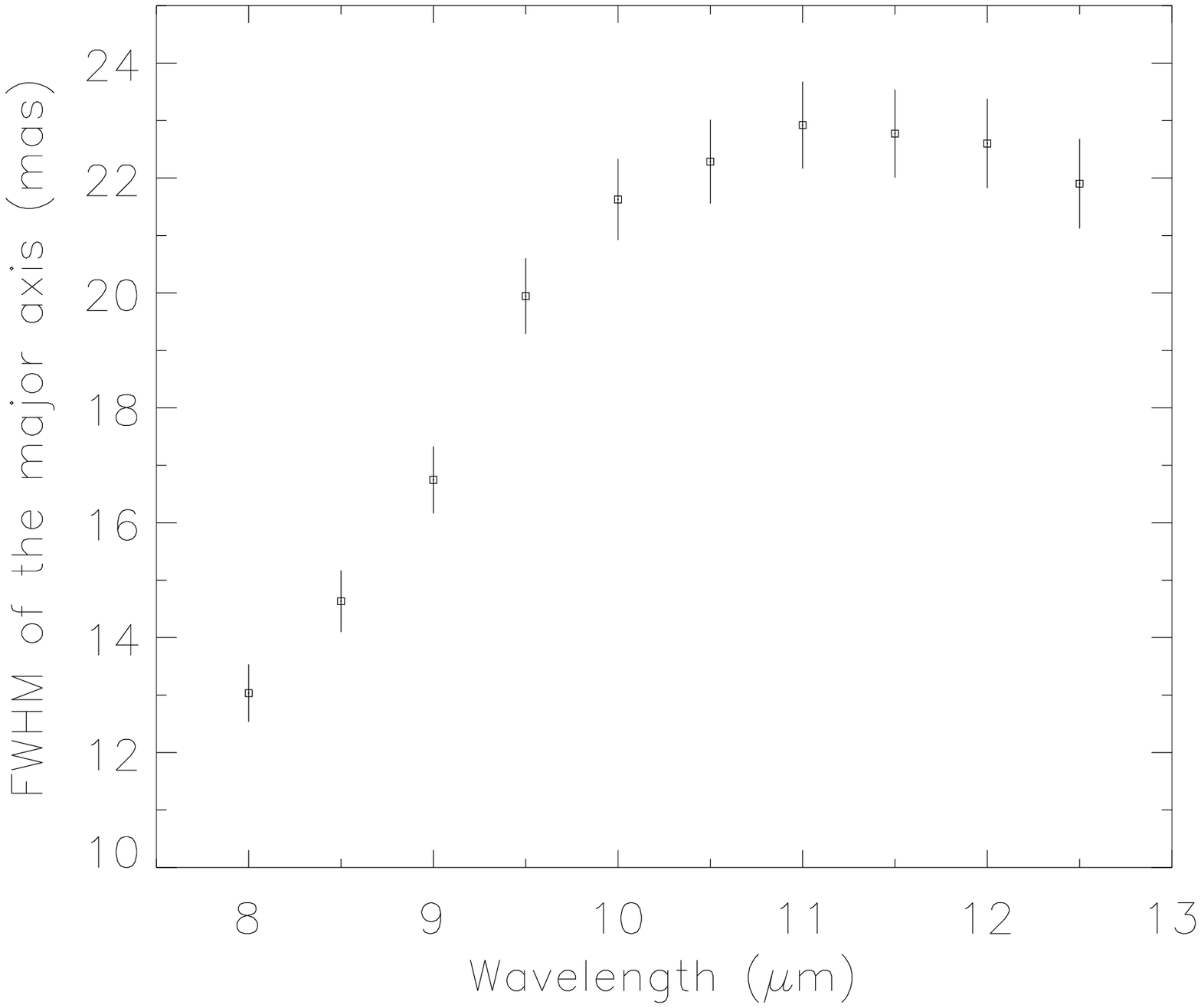}
\hspace{1.3cm}
\includegraphics[width=6.0cm]{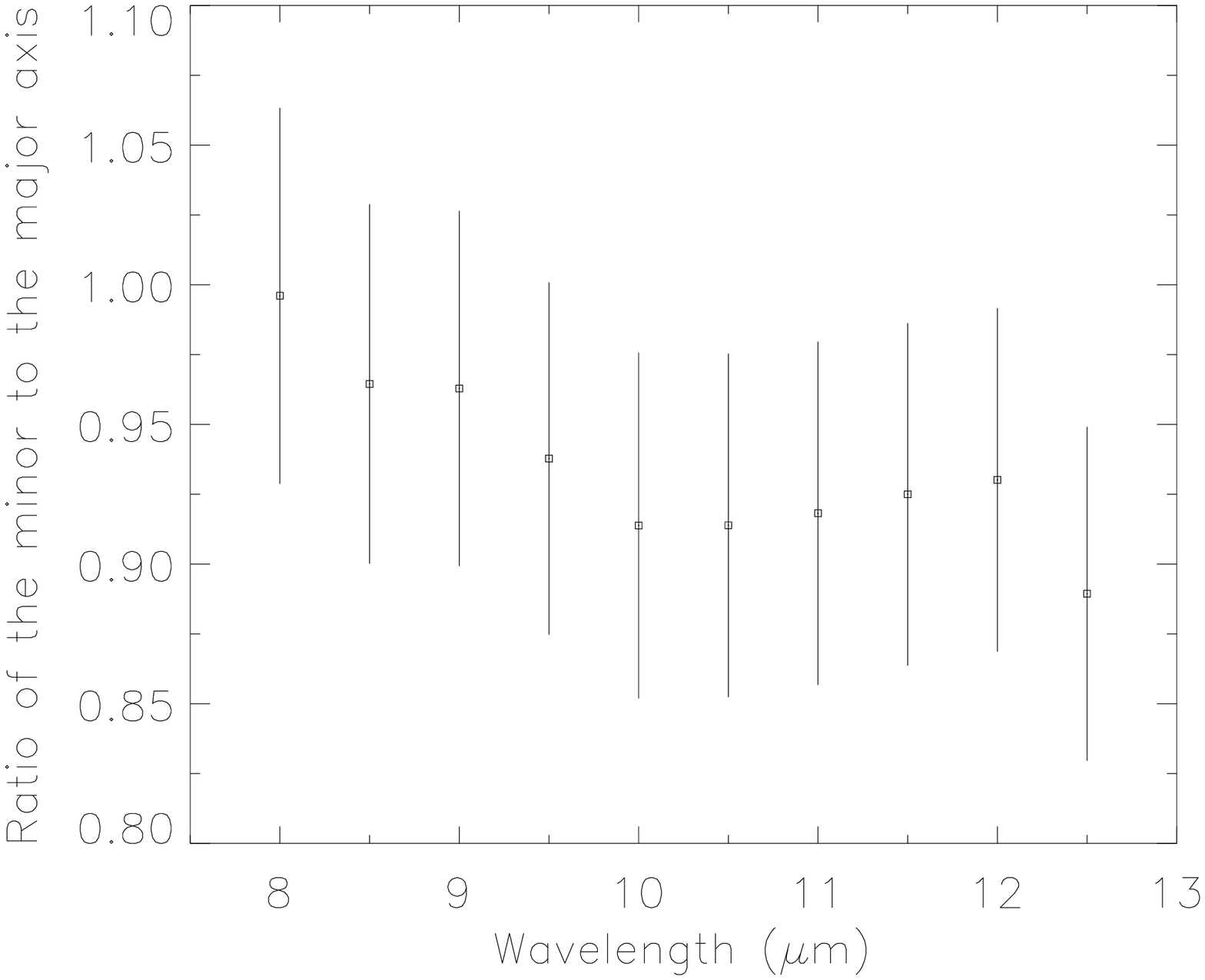}
\end{center}
\caption{\textit{Left}: full width at half maximum of the major axis for the best-fit elliptical Gaussian model as a function of wavelength. \textit{Right}: ratio of the minor to the major axis for the best-fit elliptical Gaussian model as a function of wavelength.}
\label{fig_morp_para}
\end{figure*}

By completing a non-linear least squares fitting, based on the Levenberg-Marquardt method, we obtained the best-fit elliptical Gaussian model to the data with a reduced $\chi^{2}$ of 0.27. Figure~\ref{uv_cov} shows the spectrally-dispersed $uv$ coverage of all the observations (see Table~\ref{journal}). Figure~\ref{fig_morp_uv} provides the brightness distribution of the best-fit elliptical Gaussian model of \object{HM\,Sge} between 8 and 12.5 $\mu$m. Figure~\ref{fig_morp_para} shows the resulting FWHM of the major axis, and the ratio of the minor to the major axis for the best-fit elliptical Gaussian model as a function of wavelength.

The best-fit model is obtained for an extension of the structure perpendicular to the binary axis located at PA=130$^\circ$ NE \citep{eyres01} as presented in our previous work. The length of the major axis is also approximatively the same as that published in \citet{sacuto07}. However, the level of elongation {\it increases} from 8 to 12.5 $\mu$m, in contrast to results found before. This difference can be explained by the wider $uv$ coverage allowing the geometry of the structure to be more tightly constrained.

\section{Conclusions}
\label{concl}

The wider $uv$ coverage data has allowed us to confirm that the dust envelope around the Mira is described well by an optically thick dust shell that consists predominantly of amorphous silicates. We also confirmed the longer extension of the circumstellar envelope in a direction perpendicular to the binary axis, and the increase in the size of the structure between 8 and 12.5 $\mu$m. It has also been demonstrated that the amount of elongation is higher at 12.5 $\mu$m and steadily decreases toward 8 $\mu$m, but the ratio between the two axis remains limited at most to 0.85, implying only a slight departure from spherical symmetry. This behavior can naturally be interpreted. The layers emitting most of the 8 $\mu$m flux, located closer to the photosphere, are the least affected by the White Dwarf, situated far (60 AU; \citealt{eyres01}) from the Mira. At 13 $\mu$m, the interferometer probes farther regions that are more affected by the energetic but diluted wind originating in the erupting White Dwarf. We also note that the new baselines, recorded in 2007, range from 60 to 89m whereas the previous ones ranged from 32 to 59m and probed outer regions of the dusty envelope, which were potentially more affected by the White Dwarf. Therefore, we expect the amount of elongation to be higher, as reported in \citet{sacuto07}, although the poor $uv$ coverage of the 2005-2006 observations does not allow us to estimate it accurately.

These MIDI/VLTI interferometric observations probe a compact region that includes the giant star and its vicinity to a radius of about 30 mas, set by the smallest baselines ($\sim$32m). There is a clear requirement for near and mid-IR high dynamics, imaging at the diffraction limit of a 8-10m class telescopes, to probe the inner nebula (0.8$\arcsec$ wide) already studied with the HST \citep{eyres01} and in the radio \citep{richards99}.

We are close to the limits of the imaging capabilities of the VLTI used with the two telescopes recombiner MIDI for studying such a complex source. It is also uncertain whether the three telescopes recombiner AMBER, operating in the near-IR, can provide a substantial increase in information about the morphology of the system, because the (spherical) Mira and hot dust in the inner wind still dominate. A significant improvement in the quality of image reconstruction must await the advent of the second generation instrument MATISSE \citep{lopez06}.

\begin{acknowledgements}
      
We thank C.~B. Markwardt for the available IDL routines used to perform least-squares surface fitting. We thank J. Hron, W. Nowotny, and B. Aringer for the helpful discussions. St\'ephane Sacuto acknowledges funding by the Austrian Science Fund FWF under the project P19503-N13.

\end{acknowledgements}

%
%

%
%

\end{document}